\documentclass[sigconf]{acmart}
\usepackage{subfig}
\usepackage{multirow}
\usepackage[table]{xcolor}
\usepackage{tabularx}
\usepackage{makecell}

\definecolor{rq1color}{HTML}{FADBD8} 
\definecolor{rq2color}{HTML}{D1F2EB} 
\definecolor{rq3color}{HTML}{E5E8E8} 

\usepackage{booktabs}
\usepackage{tabularx}
\usepackage{array}
\usepackage{fontawesome5}

\newcolumntype{Y}{>{\raggedright\arraybackslash}X}
\newcolumntype{P}[1]{>{\raggedright\arraybackslash}p{#1}}

\newcommand{\tech}{\faIcon{microchip}}     
\newcommand{\struct}{\faIcon{sitemap}}     
\newcommand{\econ}{\faIcon{dollar-sign}}   
\newcommand{\inter}{\faIcon{handshake}}    
\newcommand{\datafrag}{\faIcon{database}}  
\newcommand{\icons}[1]{\mbox{#1}}

\AtBeginDocument{%
  }

\setcopyright{acmlicensed}
\copyrightyear{2018}
\acmYear{2018}
\acmDOI{XXXXXXX.XXXXXXX}
\acmConference[Conference acronym 'XX]{Make sure to enter the correct
  conference title from your rights confirmation email}{June 03--05,
  2018}{Woodstock, NY}
\acmISBN{978-1-4503-XXXX-X/2018/06}

\begin{document}

\title[Generative AI-Enabled Refund Fraud in Chinese E-Commerce:\\ Investigation on Merchants and Platform Workers]{Generative AI-Enabled Refund Fraud in Chinese E-Commerce: Investigation on Merchants and Platform Workers}

\author{Shuning Zhang}
\email{zsn23@mails.tsinghua.edu.cn}
\affiliation{
    \institution{Tsinghua University}
    \city{Beijing}
    \country{China}
}

\author{Eve He}
\email{eve.he@wisc.edu}
\affiliation{
    \institution{University of Wisconsin-Madison}
    \city{Madison}
    \state{Wisconsin}
    \country{U.S.}
}

\author{Xiao Zhan}
\email{xzhan1@upv.edu.es}
\affiliation{
    \institution{Universitat Politècnica de València}
    \city{València}
    \country{Spain}
}

\author{Shijing He}
\email{shijing.he@kcl.ac.uk}
\affiliation{
    \institution{King's College London}
    \city{London}
    \country{U.K.}
}

\author{Robert Xiao}
\email{brx@cs.ubc.ca}
\affiliation{
    \institution{University of British Columbia}
    \city{Vancouver}
    \state{British Columbia}
    \country{Canada}
}

\author{Xin Yi}
\authornote{Corresponding author.}
\email{yixin@tsinghua.edu.cn}
\affiliation{
    \institution{Tsinghua University}
    \city{Beijing}
    \country{China}
}

\author{Hewu Li}
\email{lihewu@cernet.edu.cn}
\affiliation{
    \institution{Tsinghua University}
    \city{Beijing}
    \country{China}
}
\renewcommand{\shortauthors}{Trovato et al.}

\begin{abstract}
  E-commerce dispute resolution typically relies on the security assumption that digital evidence truthfully reflects physical reality. Generative AI (GenAI) invalidates this threat model, enabling attackers to fabricate hyper-realistic evidence of product defects at negligible cost. Through semi-structured interviews with merchants (N=17) and platform workers (N=13) in the Chinese e-commerce market, we characterize this shift toward GenAI-enabled scalable fabrication. We outline a taxonomy of four GenAI-enabled threat vectors across the transaction, dispute, logistics and communication phases, highlighting how attackers exploit GenAI to synthesize physically plausible product defects at scale. To mitigate these threats, platforms and merchants are adapting verification strategies, relying on AI tools for automated screening and adversarial interrogation (e.g., requesting multi-angle videos) to increase attack complexity. However, we find several challenges that hinder the adoption of these defenses, including implementation hurdles like structural platform constraints and fundamental limitations regarding the technical sophistication of GenAI. We conclude by outlining design implications for privacy-preserving cross-platform fraud databases, and traceability mechanisms such as embedding verifiable material anchors into the product.
\end{abstract}


\keywords{Do, Not, Use, This, Code, Put, the, Correct, Terms, for,
  Your, Paper}


\maketitle

\section{Introduction}
E-commerce has evolved into a critical pillar of the modern retail economy. In China, major platforms, including Taobao, Douyin, and Pinduoduo, facilitate immense transaction volumes: for example, they collectively generated 1.7 trillion RMB in sales during the single-day 2025 ``Double 11'' shopping festival. The scale of these transactions is accompanied by a large number of return requests, refund claims, and buyer-seller disputes, placing substantial pressure on platform adjudication systems. To maintain efficiency and consumer trust, platforms have adopted automated dispute resolution workflows and refund policies. These systems operate on the security assumption that digital evidence provided by buyers, typically photographs of defective products, accurately reflects items' physical conditions~\cite{patton2004technologies}. Within these models, refund requests are automatically approved if they satisfy specific criteria, such as price thresholds and specific keywords.

However, this threat model is currently failing. The rapid evolution of Generative AI (GenAI) is altering the economics of refund fraud~\cite{hu2025vision,hagos2024recent}. News reports that approximately 50\% of appeals from merchants concern inappropriate refund issues\footnote{https://www.pai.com.cn/p/01keqvq9m5tee6dzmgd122m6tx}. Attackers now utilize widely available GenAI tools to fabricate evidence of non-existent defects, such as mold on fresh food or cracks on electronic screens, to seek illegitimate refunds\footnote{https://www.news.cn/legal/20251126/dd9821834d114ce99c82ce79cfaa7185/c.html}. This problem also extends beyond China: global fraud reports indicate that refund and policy abuse have been rising internationally, with 57\% of surveyed merchants reporting increased rates over the past year\footnote{https://merchantriskcouncil.org/learning/mrc-exclusive-reports/global-payments-and-fraud-report/2025-global-payments-and-fraud-report}.


Despite the severity of this threat, the specific mechanisms of this refund fraud\footnote{also called reverse logistics refund, return fraud.} remain underexplored. Prior literature has largely examined e-commerce fraud through social or economic lenses~\cite{zhang2013trust}, developed detection algorithms~\cite{mutemi2024commerce,zhang2022efraudcom}, or focused on deepfake detection algorithms in isolation~\cite{tang2024deepmark,joslin2024double}. They overlook how GenAI techniques are operationalized within real-world e-commerce refund workflows. It remains unclear \textit{how} these adversarial inputs bypass existing verification, and \textit{what} barriers prevent effective mitigation.

We place special emphasis on ``Refund Only'' cases, as distinct from traditional Return to Origin (RTO) models. While RTO mandates physical verification, Refund Only prioritizes efficiency by waiving returns, especially for low-value or perishable items. This creates a critical vulnerability: GenAI-enabled fraud exploits these digital-only pathways, using fabricated evidence to enable scalable abuse. Consequently, we address the following research questions (RQs):

$\bullet$ \textbf{RQ1:} What are the threat vectors in e-commerce refund fraud, and how does GenAI change these adversarial practices?

$\bullet$ \textbf{RQ2:} What verification and resolution mechanisms do merchants and platform workers adopt?

$\bullet$ \textbf{RQ3:} What constraints prevent effective mitigation of these vulnerabilities in refund fraud?

Towards these RQs, we conducted semi-structured interviews with e-commerce merchants (N=17) and platform workers (N=13, 1 overlap with merchants) who were involved in refund frauds. 

Regarding RQ1, we categorized refund fraud attack vectors, especially the GenAI-enabled ones, spanning the pre-shipment, dispute, logistics and communication phases. We find that GenAI facilitates scalable fabrication by producing hyper-realistic, physically plausible defects that deceive human heuristic judgment. This shift transforms refund fraud from isolated opportunistic acts into systematic, organized operations that exploit the e-commerce lifecycle.
 
Regarding RQ2, we found merchants employ multimodal verification by combining manual heuristics with general AI tools, though these tools are increasingly ineffective with rapid AI evolution. They also requested contextual evidence, but are often refused by customers. Although intervention criteria are traditionally determined by factors like price and customer scale, the negligible cost of GenAI-enabled fraud shifts these criteria, forcing merchants to internalizes losses. They negotiate with customers or escalate disputes to platforms, however find it difficult to succeed with a lack of grounded AI-edited evidence. 


Regarding RQ3, we revealed that merchants and platform workers faced three constraints: technical and infrastructural, procedural and interactional, and economic and resource constraints. They lacked dedicated tools to distinguish and verify AI-edited artifacts, and their own capabilities are also limited in the AI era. The negligible cost of GenAI-enabled fraud makes verification hard to scale concurrently, where the cost of proof exceeds the product's value. Merchants even lacked ground truth, particularly for perishable goods, which prevent them from refuting claims. These are further complicated by platforms' consumer-centric bias, which shifts the burden of proof onto merchants. To address these vulnerabilities, we advocate for cross-platform reputation sharing, physical-digital grounding through material anchors, and risk adaptive adjudication workflows.

Collectively, this paper makes the following contributions:

$\bullet$ We categorize refund fraud attack vectors in e-commerce contexts, especially GenAI-enabled ones, across four stages of the transaction lifecycle.

$\bullet$ We provide insights into the refund fraud vulnerabilities within e-commerce systems, highlighting how the negligible cost of GenAI-enabled fraud shifts the burden of proof.

$\bullet$ We outline implications for GenAI-enabled fraud resolution, highlighting collaborative database sharing, physical-digital grounding, and risk adaptive adjudication workflows.

\section{Background: E-Commerce Ecosystem and Flow}

The Chinese e-commerce landscape is characterized by business models differing in scale and norms from Western counterparts. Major players include Alibaba/Taobao~\cite{taobao2025} and Jingdong~\cite{jd2025}, which primarily focus on traditional search-based retail, as well as Douyin~\cite{douyinec2025} and Pinduoduo~\cite{pinduoduo2025}, which emphasize social commerce through live-streaming and group-buying mechanisms~\cite{wang2022live,ji2025motivates}. Most platforms operate on a tripartite model involving the platform, merchants and consumers, as in Figure~\ref{fig:ecosystem}. Platforms typically generate revenue by charging commissions on successful transactions. The typical flow proceeds through the following phases:

\begin{figure}[!htbp]
    \centering
    \includegraphics[width=0.5\textwidth]{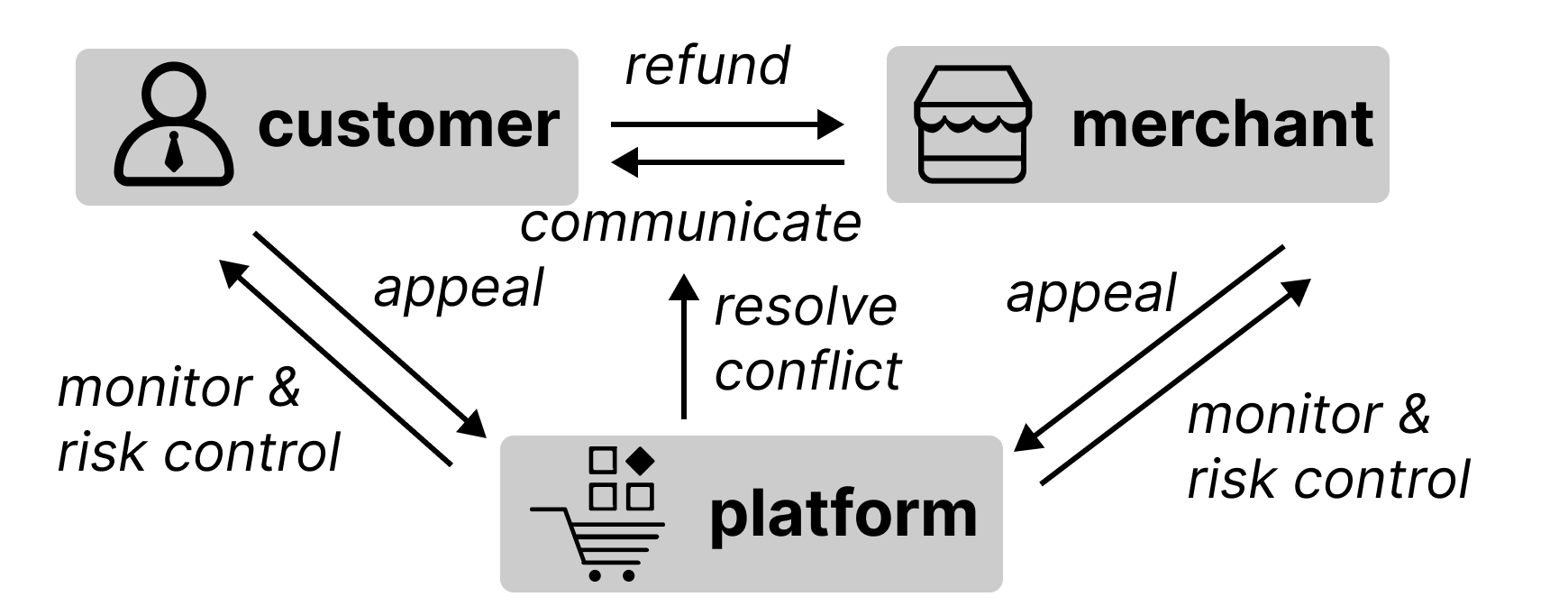}
    \caption{The ecosystem around refund fraud, involving customers, platforms and merchants.}
    \label{fig:ecosystem}
\end{figure}

\textbf{\textit{Pre-transaction.}} Unlike the centralized retail models common in North America (e.g., Amazon), the Chinese e-commerce landscape (e.g., Taobao, Pinduoduo, Douyin) operates primarily on a decentralized, storefront-centric model. While western platforms typically enforce uniform return policies and handle logistics directly, Chinese merchants have greater autonomy to manage their storefronts, though they remain bound by platform-imposed metrics. In this phase, attackers may abuse inappropriate discounts or use multiple accounts to repeatedly obtain discounts.

\textbf{\textit{Fulfillment.}} Following payment, the product is shipped to the user. Transactions in China are overwhelmingly executed via bank debit or mobile wallet transfers paid directly to the platform, rather than credit. This means that disputes must be resolved with the payee directly, typically the platform. Platforms thus enforce delivery checks together with merchants and customers.

\textbf{\textit{Refund request.}} Consumers dissatisfied with a product typically initiate a refund request via the platform's communication channel. Due to the labor-intensive nature of after-sales support, platforms and merchants manage these requests using a combination of in-house teams, third-party services, and AI chatbots. These communication channels are concurrently monitored by the platform. If a consumer's messages trigger specific keyword thresholds, the platform's automated systems may proactively intervene, review, and (often) approve the refund. Otherwise, merchants can interactively request evidence to support the claim. Approved requests result in direct refunds, while denied ones can be escalated by customers to platform mediation. However, this reliance on chatbots for dispute mediation also leads to high volumes of misclassification.

An important feature of the Chinese ecosystem is the refund-only policy, a mechanism to optimize efficiency and build customer trust by allowing customers to receive refunds without returning the physical item. For platforms, this automates dispute resolution and reduces administrative overhead. For merchants, it theoretically reduces management costs by waiving the expensive reverse logistics (Return-to-Origin, RtO) typically required for products, especially low-value or perishable ones~\cite{martinez2022using}. However, merchants are forced to internalize the financial losses of this policy, while lacking the agency to opt out\footnote{https://daxueconsulting.com/consumer-rights-regulations-in-china/}. ``Refund-fraud-as-a-service'' software is even widely distributed via messaging platforms~\cite{George2024RefundAbuse,Perry2026ImageBasedFraud}.

As e-commerce becomes more ubiquitous, global retailers like Amazon are also introducing similar refund-only policies to reduce costs, making them vulnerable to similar attacks.

\textbf{\textit{Adjudication.}} Disputes that fail automated mediation escalate to human reviewers. Owing to the time and effort required to review evidence, merchants and platforms favor rapid dispute resolution over strict evidence verification. Users may also pressure merchants into a favorable outcome with the threat of negative social media attention or reports to authorities.


In refund processes, the main dispute resolution process only involves merchants and customers. Platforms may proactively intervene if some of the keywords signaled conflict escalation. In these cases, they mostly favor the customers. In other cases, platforms intervened when these two parties disagree on disputes, and either customer or merchant can initiate a review by the platform.

\textbf{\textit{Product return and transaction reversal.}} In a refund-only request, the payment is directly returned to the customer. Otherwise, for RtO, the customer must ship the product back to the merchant. Platforms determine the timing of the refund based on the shipping process and customers' trustworthiness. 

\section{Related Work}

\subsection{Online Fraud}

Existing literature on online fraud encompasses general scam domains, digital advertising manipulation, and e-commerce misconduct. Research on general scams primarily focus on characterizing malicious web infrastructure, quantifying user exposure to fraudulent domains~\cite{kotzias2023scamdog,kotzias2025ctrl+,bitaab2023beyond}, and analyzing search engine exploitation alongside consumer complaints to expose deceptive practices~\cite{liu2025nokescam,ma2023investigating}. Similarly, investigations into advertising fraud address the manipulation of ad impressions and search traffic~\cite{deblasio2017exploring,springborn2013impression,sun2021understanding}, noting that adversaries increasingly deploy device emulators and humanoid algorithms to evade behavioral detection~\cite{wen2018fraus,zhu2021dissecting}.

Within the specific domain of e-commerce, scholars have documented a diverse taxonomy of fraudulent activities~\cite{badotra2021systematic}. This includes financial exploits such as identity theft, card-not-present fraud using stolen credentials~\cite{abi2020boxer}, and friendly fraud, where customers illegitimately request bank chargebacks after receiving their purchases~\cite{o2025impact}. Other deceptive practices target platform operations and market dynamics. For instance, abusive dropshipping involves the unauthorized listing of third-party inventory at inflated prices, fulfilling orders by purchasing directly from the original merchant using the victim's shipping details~\cite{arunasalam2024dark}. Similarly, adversaries deploy triangulation schemes through counterfeit storefronts and execute network traffic manipulation to artificially simulate customer engagement~\cite{carpineto2017learning,kotzias2023scamdog}.

Beyond transaction scams, e-commerce integrity is compromised by platform vulnerabilities and reputation manipulation. Attackers exploit logic flaws to bypass payment~\cite{pagey2023all}. Furthermore, underground economics distort market trust through self-reputation escalation via fabricated transactions~\cite{xu2017empirical} and the orchestration of incentivized reviews for cash-out services~\cite{oak2025towards,van2018plug}. To counter these threats, defensive frameworks primarily leverage time-dependent behavioral modeling to detect malicious usage patterns~\cite{weng2019cats,chen2025identifying}. While those literature covers automated traffic manipulation, payment exploits and behavioral modeling, we focus on the use of GenAI to fabricate high-fidelity visual evidence for refund fraud.

\subsection{AI Risks, Deception, and Manipulation}

The rapid evolution of GenAI significantly expanded the landscape of digital deception, ranging from safety threats~\cite{park2024ai} to the creation of hyper-realistic synthetic profiles~\cite{mink2022deepphish,joslin2024double,kolomeets2024face,zhang2024confrontation}. While technical efforts have focused on developing unified detection frameworks~\cite{li2021deepfake} or evaluating the efficacy of state-of-the-art detectors across various modalities~\cite{yan2025voicewukong,cho2023towards}, human observers continue to struggle with high error rates when identifying AI-generated disinformation~\cite{mink2024s}. This struggle is exacerbated by a reliance on fallible linguistic cues and intuition rather than robust forensic signals~\cite{warren2024better,kharvi2024understanding}.

Deceptive manipulation also extends across the technical pipeline, from data acquisition to the user interface. Recent literature documented signal-level attacks that compromise Image Signal Processing (ISP) integrity via data poisoning~\cite{zhu2025neural} and presentation-layer audits designed to detect deceptive patterns~\cite{nayak2025automatically}. At the policy level, researchers examined the constraints of content moderation within GenAI ecosystems and how users navigate these boundaries~\cite{gao2025cannot}. Different from these studies, our work investigates practitioner practices towards frauds in e-commerce contexts.

\subsection{Security and Integrity in E-Commerce}

Ensuring security and integrity in e-commerce requires addressing both AI-induced vulnerabilities and fraudulent ecosystems. Existing papers design algorithms to mitigate AI-induced risks, empirically analyze malicious underground economies and develop behavior-modeling systems.

Regarding content integrity, the proliferation of generative AI introduces novel risks alongside new defensive capabilities. Recent studies have established benchmarks to identify evasive content and AI-generated risks~\cite{xu2025evade,zhang2025aiguard}, differentiated the characteristics of AI-generated fake reviews from human-crafted ones~\cite{zhao2025ai}, and explored the ethical and practical efficacy of deploying generative AI for real-time fraud detection~\cite{tyagi2025generative}.

Beyond AI-specific threats, substantial work empirically investigates the operational mechanics of e-commerce cybercrime. Researchers have characterized scam websites~\cite{kotzias2023scamdog}, monitored illicit underground communities and messaging networks~\cite{arunasalam2025characterizing,wang2020into}, and analyzed widespread misconduct via large-scale customer complaints~\cite{ma2023investigating}. Specific fraudulent practices have also been thoroughly audited, providing insights into reshipping scams~\cite{hao2015drops}, affiliate cookie-stuffing~\cite{chachra2015affiliate,snyder2016characterizing}, seller-reputation escalation~\cite{xu2015commerce}, and concession-abuse-as-a-service~\cite{sun2021having}.

To counter these threats, prior studies rely on advancing machine learning frameworks for behavioral modeling. This includes leveraging sequence encoders, graph neural networks, and process mining to capture complex, time-dependent, cross-domain, or group-based fraudulent behaviors~\cite{chen2025identifying,zhu2020modeling,yu2024temporal,zhang2022efraudcom,ling2023learned,kumar2024multi}. Other technical defensive strategies focus on enhancing data representation through knowledge graphs~\cite{wang2020representing} and enabling privacy-preserving fraud intelligence sharing among merchants~\cite{arp2018privacy}. While these literature emphasizes algorithmic detection and automated behavioral modeling, we construct a taxonomy of GenAI-enabled threat vectors, and expose the vulnerabilities that shift the burden of proof.

\section{Methodology}

To provide a comprehensive understanding of refund fraud, we designed semi-structured interviews for two key groups: merchants and platform workers. Merchants are scoped as personnel responsible for after-sales services within individual storefronts, while platform workers refer to individuals managing after-sales operations within the e-commerce platform.

\subsection{Participants and Apparatus}

To synthesize a multi-faceted perspective, we recruited 29 participants, comprising 17 individuals with merchant experience (7 males, 10 females) and 13 with platform-side experience (3 males, 10 females). One participant possessed professional experience in both domains. Merchant-side participants covered diverse product categories, including food, cosmetics, apparel, and general goods. These participants operated across several major platforms, most frequently Taobao (9), Pinduoduo (7), and Douyin (4). Platform-side participants had experience in Pinduoduo (7), Taobao (2), Meituan (1), and Dewu (2), with roles including customer service, after-sales support, process optimization, and technical operations. Each participant was compensated 100RMB according to local wage standards. The interview was approved by our university's Institutional Review Board (IRB).

\subsection{Interview Design and Procedure}

We structured interview scripts across merchants and platform workers progressively, moving from individual case observations to systemic policy analysis, to directly address our RQs regarding the identification of malicious attackers' attack vectors, their practices, and the limitations of current defenses.

Prior to the interviews, we obtained informed consent from all merchants and platform workers, with a detailed explanation of their rights to withdraw from the interview or request data deletion without justification. We first asked about participants' professional experiences. After introducing their roles and general workflows, they were asked to share specific cases of refund disputes, particularly those involving AI-enabled or synthetic evidence. For each case, we prompted participants to describe their reactions and identification processes, including, for example whether they could distinguish fraudulent content, specific tools or indicators they relied upon, and the reasoning behind their assessments. We then prompted participants' decision-making logic and operational practices. Finally, participants shared the systemic challenges they face and their expectations for future countermeasures. Each session concluded with an open-ended question for additional thoughts. During the interview, we added follow-up questions when participants provided interesting responses to previous questions. The interview scripts for both groups are available in Appendix~\ref{app:interview_script}.

All interviews were conducted through Tencent Meeting with a institutional account, and were audio-recorded, with an average length of 48.3 minutes (SD=13.1 minutes, range=25-82 minutes). Transcripts were automatically generated from recordings using Tencent Meeting's built-in tools and checked by one primary author.

\subsection{Data Analysis}

We performed thematic analysis~\cite{braun2006using} on the transcribed text of all interviews. To leverage their complementary perspectives, we integrated data from both groups into a unified coding process. Two primary authors first coded 4 scripts (2 from the merchants and 2 from the platform) to set the initial codebook. Then these two authors discussed the codebook to resolve potential disagreements. After that, these two authors separately coded the remaining interview scripts, with intermittent discussions to resolve conflicts. They discussed and synthesized themes and sub-themes after having coded all scripts, forming the final codebook (see Appendix~\ref{sec:codebook}). Due to the inductive nature of these studies and the issues of calculating inter-rater reliability in such studies~\cite{mcdonald2019reliability}, we opted to not calculate inter-rater reliability. One author translated Chinese materials (including codes, themes and quotes) into English, and the rest of the author team checked the correctness of the codes. We counted appearance frequencies post-coding.

\section{RQ1: Threat Vectors Across Refund Lifecycles} 

Inspired by e-commerce lifecycle categorization~\footnote{}, we categorized threat vectors by their occurrence in the transaction lifecycle, as shown in Table~\ref{tab:mapping_table}. Below, we use [G] to indicate factors that are enabled or amplified by GenAI tools.


\begin{table*}[h]
\centering
\caption{Refund fraud threat vectors, verification lifecycle, and constraints. Note: [G] denotes threat factors enabled or amplified by GenAI. Color coding represents the research framework, \colorbox{rq1color}{RQ1: Threat}, \colorbox{rq2color}{RQ2: Practices}, and \colorbox{rq3color}{RQ3: Challenges}.}
\label{tab:mapping_table}
\footnotesize
\begin{tabularx}{\textwidth}{p{2.2cm}XXXXX}
\hline
\rowcolor{gray!20}
\textbf{Lifecycle Phase} & \cellcolor{rq1color}\textbf{Attack Model (RQ1)} & \cellcolor{rq2color}\textbf{Verification (RQ2)} & \cellcolor{rq2color}\textbf{Intervention (RQ2)} & \cellcolor{rq2color}\textbf{Resolution (RQ2)} & \cellcolor{rq3color}\textbf{Constraints (RQ3)} \\ \hline

\textbf{1. Pre-shipment \& Transaction} & 
$\bullet$ Automated arbitrage \newline 
$\bullet$ [G] Synthetic persona \newline 
$\bullet$ Entry barrier reduction & 
$\bullet$ Algorithmic risk screening \newline 
$\bullet$ Reputation-based verification & 
$\bullet$ Automated platform triage \newline 
$\bullet$ Account history filtering & 
$\bullet$ Data-anchored reporting \newline 
$\bullet$ Batch behavior flagging & 
$\bullet$ AI outpaces detection \newline 
$\bullet$ Data fragmentation \\ \hline

\textbf{2. Dispute \& Evidence Submission} & 
$\bullet$ [G] Hyper-realistic defect fabrication \newline 
$\bullet$ [G] Contextual forgery & 
$\bullet$ Physics-consistency checks \newline 
$\bullet$ Spatial interrogation \newline 
$\bullet$ Digital forensics (Exif) & 
$\bullet$ Refund-only exception handling \newline 
$\bullet$ Physical plausibility gateways & 
$\bullet$ Supervisory escalation \newline 
$\bullet$ Multi-format re-verification & 
$\bullet$ Asymmetric cost-of-proof \newline 
$\bullet$ Missing ground truth \\ \hline

\textbf{3. Logistics \& Re-shipment} & 
$\bullet$ [G] Asynchronous substitution \newline 
$\bullet$ Carrier interception & 
$\bullet$ Logic verification \newline 
$\bullet$ Physical material anchors & 
$\bullet$ Store scale capacity limits \newline 
$\bullet$ Procedural heterogeneity checks & 
$\bullet$ Formal video appeals \newline 
$\bullet$ Mandatory return/exchange & 
$\bullet$ Buyer-focused bias \newline 
$\bullet$ Admin irreversibility \\ \hline

\textbf{4. Communication \& Interaction} & 
$\bullet$ Off-platform double extortion \newline 
$\bullet$ Regulatory exploitation & 
$\bullet$ Historical chat trace audits \newline 
$\bullet$ Reputation-based tagging & 
$\bullet$ Algorithmic loss thresholds \newline 
$\bullet$ Risk-weighted decision & 
$\bullet$ Strategic loss absorption \newline 
$\bullet$ Selective disengagement & 
$\bullet$ Automation bias \newline 
$\bullet$ Suppressed merchant discourse \\ \hline
\end{tabularx}
\end{table*}

\subsection{Pre-shipment and Transaction Phase}

While legitimate users in this phase use promotional incentives for consumption at a retail scale, attackers exploit platform promotions, coupons and shipping subsidies to acquire large volumes of discounted goods. Adversaries leverage GenAI and other technical tools to scale these operations.

\subsubsection{Automated promotion and subsidy arbitrage}

Participants noted that attackers operate as coordinated professional groups using specific technical tools to exploit platform logic. For instance, adversaries use backend data extraction software to bypass coupon demographic restrictions (P1) and acquire gray-market phone numbers to register multiple fraudulent accounts (P6). To exploit threshold-based promotional algorithms, attackers bundle items to trigger aggregate discounts, subsequently refunding the majority of the order to retain specific goods at reduced costs (M2). These traffic surges rapidly deplete merchant inventory (P2), generating massive illicit profits (M14). Furthermore, these groups maintain dedicated warehouses exclusively to process high-volume returns, systematically extracting platform-provided shipping subsidies (P1).

\subsubsection{[G] Synthetic persona construction}

Attackers use GenAI to fabricate human-like behavioral histories and profiles to evade detection algorithms designed to identify robotic patterns. This synthetic credibility allows coordinated professional groups, using gray-market accounts and backend data extraction, to scale promotional and subsidy frauds.

\subsubsection{[G] Entry barrier reduction} 

GenAI tools lower the technical skills required for fabrication, allowing a broader demographic to engage in systematic fraud. As M13 observed, \textit{``after AI tools are used, everyone can just give an instruction ... the user base expands, the cost of use drops, and more people get involved.''}

\subsection{Dispute and Evidence Submission Phase}

While legitimate consumers in this phase provide authentic evidence of verified product defects, attackers rely on GenAI or other software to fabricate visual and contextual evidence, obtaining refunds without returning physical items.

\subsubsection{[G] Hyper-realistic defect fabrication}

Adversaries use GenAI to synthesize defects that respect material properties and physical logic, making artifacts harder to detect via manual heuristics. Examples include generating ``moldy spots on fresh produce like durians'' (P11), ``synthetic cracks on shoe edges'' (P10), or ``snapped plastic phone cases'' (M2). 

Merchants indicated that adversaries leverage GenAI and editing software to create adversarial visual inputs that deceive both automated classifiers and human auditors. Participants observed that attackers targeted expensive or perishable goods to secure refunds without returning the physical items. For example, attackers digitally alter images to show snapped phone cases (M2), generate synthetic cracks on shoe edges (P10), or manipulate pure red items into pink versions to claim the product was not as described (P5). P11 also highlighted the abuse of food hygiene policies, noting that attackers use AI to generate moldy spots on fresh produce like durians. When digital tools are insufficient, merchants reported that attackers physically manipulate goods. P6 observed attackers ripping holes in clothing, and M10 noted instances of intentionally breaking mirrors. M11 documented this physical destruction, stating that when the buyer took the photo they only manipulated the necklace's broken part, but other intact parts could be reused. Furthermore, attackers synthesize hard-to-verify flaws like liquid leakage to request compensation from merchants (M8).

\subsubsection{[G] Contextual attribute forgery and digital recycling}

\begin{figure}[!htbp]
    \subfloat[]{
        \includegraphics[height=0.14\textwidth]{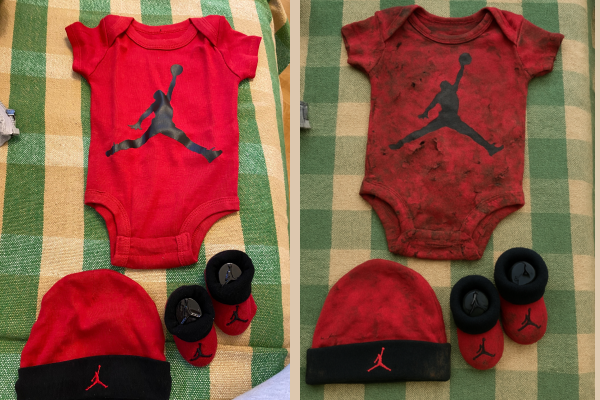}
    }
    \subfloat[]{
        \includegraphics[height=0.14\textwidth]{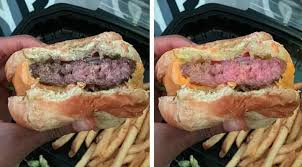}
    }
    \caption{Illustrations of manipulated images used for refund fraud. The right side of each image is the manipulated one.}
\end{figure}

Beyond product images, GenAI facilitates contextual synthesis to forge a complete evidentiary chain. P3 reported that adversaries \textit{``forge external communication records to falsely claim merchant agreements to refunds''}. This allows for ``digital recycling,'' where a single manipulated photo is used to \textit{``claim compensation from four different shops concurrently.''} (M3)

Participants detailed how attackers synthesize contextual administrative evidence and reuse adversarial assets across multiple domains to enhance the plausibility of their claims. P3 reported that adversaries forge external communication records to falsely claim that merchants agreed to refunds. Merchants also suffered from repeated submissions of the same fabricated image. M3 found a single manipulated photo was used to claim compensation from four different shops concurrently. To enforce compliance and bypass algorithmic visibility protections, attackers weaponize this fabricated content on public forums. For instance, M12 stated that even though merchants politely explain that the images are inaccurate and do not represent the products, buyers threaten to post the content as a negative review. Merchants noted this behavior inflicts reputational damage, leading to deductions in store ratings and reduced search exposure even when the content is tagged as AI-generated (M12).

\subsection{Logistics and Re-shipment Phase}

Even in RtO processes, attackers can exploit the gaps between platform adjudication and product re-shipping, causing a result similar to refund-only fraud.

\subsubsection{[G] Algorithmic bypass and asynchronous substitution}

Adversaries target platforms' dispute-resolution chatbots that rely on rudimentary visual heuristics, which frequently misclassify \textit{``an AI-generated image as a quality issue''} (P8). By presenting GenAI-fabricated evidence, attackers manipulate automated logic to secure financial restitution before the merchant can physically inspect the return. As M14 observed, \textit{``the system automatically determines quality issues based on simple descriptions and immediately issues a refund.''} This asynchronous approval enables fraudulent product substitution, where attackers return counterfeit goods (M1, M9) or initiate refunds for items still in transit. Even when merchants later uncover the discrepancy and present forensic evidence, algorithmic bias towards customers persists: platforms often retain the original decisions for a refund. Platforms often grant the refund before the merchant has even received the returned goods (M10), or mandate a resolution where merchants must still bear shipping costs (P8). In this stage, participants also mentioned problems with product swap or wardrobing (the return of worn or used goods).

\subsubsection{Carrier-side interception}
Merchants reported that attackers exploit specific timing vulnerabilities to target high-value assets. Once the platform prematurely releases funds, attackers initiate a return shipment and immediately request a carrier interception to reroute the package back to them. This tactic allows them to retain both the item and refunded money, forcing merchants into continuous, manual monitoring of shipment trajectories to mitigate asset loss (M10).

\subsection{Communication and Interaction Phase}

\subsubsection{Off-platform double extortion and entrapment}

As reported by merchants, attackers claimed to publish manipulated images on public forums to coerce merchants into granting refunds (P12). Attackers may also try to migrate interactions onto unmonitored channels such as WeChat, demanding direct transfers before claiming a second refund on the official platform. M1 characterized the platform's inability to synchronize these external transactions as a critical system flaw. Merchants also reported that attackers entice them into these external chats, then report them for inducing off-platform contact, triggering financial penalties from platforms against merchants (M8).

\subsubsection{Regulatory exploitation and psychological coercion}

Adversaries compel settlements by escalating disputes to administrative bodies or intensifying emotional pressure on merchants. M8 observed that attackers \textit{``engage in keyword hunting to identify prohibited terms in product descriptions, filing regulatory complaints and offering to withdraw them only upon financial settlement''}. Merchants noted that attackers exploit personal data from shipping labels to subject staff to direct verbal abuse (M1). When standard synthetic forgery fails to trigger an automated refund, adversaries may even escalate to extreme psychological coercion such as suicide threats (P3). Furthermore, attackers add external pressure to merchants by threatening reports to the \textit{Bureau of Industry and Commerce} (M2). Faced with retaliatory mass ordering (P1) and biases favoring unverified allegations (M5), merchants often compromise to mitigate potential fallout (M3). M14 summarized that \textit{``it is more economical to satisfy the extortionist given the disproportionate cost of resistance, as GenAI allows attackers to scale convincing fabrications''}.

\section{RQ2: Verification Mechanisms and Resolution Strategies}

This section examines the verification practices along three aspects: multimodal verification approaches, intervention thresholds, and the resolution strategies, as shown in Figure~\ref{fig:ecosystem}.


\subsection{Multimodal Verification Approaches}

We detail the technical and human-initiated methods used to authenticate transactions and detect GenAI-enabled anomalies.

\subsubsection{Manual inspection and heuristic judgment} 

\textbf{\textit{Physics consistency checks.}} Merchants leverage domain-specific knowledge to identify visual anomalies that violate material properties or manufacturing logic. They scrutinize digital evidence for mechanical impossibilities, such as unnatural fracture patterns or structural failures that contradict product assembly standards. M2 highlighted the lack of physical plausibility in GenAI output, noting a case where \textit{``a plastic phone case suddenly snapping neatly in the middle ... makes one feel it isn't quite reasonable.''} Similarly, M5 applied technical knowledge of garment construction to refute a claim about a detached zipper, stating that because the component \textit{``is completely embedded ... [it is] impossible for a new product to have that fall off.''} These heuristics function as a filter, rejecting claims where the depicted damage defies physical laws or production constraints. 

\textbf{\textit{Adversarial spatial interrogation.}} Operators demand multi-view evidence to expose generative model's spatial inconsistency failure cases. Since current GenAI models often struggle to maintain coherent 3D geometry across different perspectives, merchants request supplementary angles of the defect to reveal rendering errors. P3 observed that \textit{``AI generation ... cannot completely replicate one-to-one [details] ... [because] it lacks a macro perspective,''} resulting in contradictions when the object is viewed from different sides. 

\textbf{\textit{Collaborative consensus.}} Verification often involves collaborative peer review, especially for challenging or high-value disputes. Initial suspicion by the individual worker serves as a start for multiple workers' cross validation, which mitigates the risk of individual bias or error. P3 described this workflow, \textit{``I would ask the colleagues around me, and they would also sense that something is indeed a bit wrong.''} In high-stakes scenarios, such as those involving luxury goods, this becomes a standardized practice, with structured deliberation involving multiple domain experts. P1 detailed this protocol: \textit{``If the amount involved is large ... I will include two colleagues ... the three of us will have a discussion ... before making the final decision.''} 

\textbf{\textit{Decentralized merchant verification.}} Verification extends beyond the platform to decentralized merchant networks that identify cross-store patterns and repeat offenders. Merchants share data within social media groups to surface recognizable patterns to peers.
P8 noted this practice of external validation, stating that they might \textit{``post it in our group for everyone to see to confirm whether a specific defect is AI-generated''}. This collective vigilance fosters them to detect anomalies, especially those identical backgrounds or reused defect images, that individual stores might miss. As P1 emphasized, \textit{``It is only during the final summary that we discover there are too many identical flaw points.''} 

\subsubsection{Technical and Automated Authentication} 

\textbf{\textit{Algorithmic risk screening.}} Merchants and platform workers also rely heavily on automated detection systems to identify behavioral and visual anomalies, creating an adversarial dynamic often described as AI vs. AI. They indicated that \textit{``manual review is increasingly preceded by algorithmic filters, as we can only use AI to identify AI''} (M2). These systems operate by detecting specific irregularity patterns, ranging from behavioral signals like \textit{``short-term bulk ordering with multiple accounts,''} (P1) to visual artifacts where physical logic is violated, such as \textit{``the edges of the eggshell cracks were very regular ... it looked a bit like it was edited by AI''} (P7). This technical assessment is usually integrated directly into adjudication logic, where risk-control tools screen for malicious batch refund fraud (P4). However, this automation functions primarily as a preliminary filter, instead of ground truth. Due to imperfect accuracy, workers manually screen algorithmic flags before communicating to customers (P5).

\textbf{\textit{Digital forensic inspection.}} Merchants and platform workers use specialized software to inspect image structure and metadata. At the visual layer, they zoom in to view suspect evidence and scrutinize digital artifacts in professional editing tools. Illustrating this strategy, M8 explained, \textit{``we currently put this picture into Photoshop ourselves to check ... whether the overall integrity is complete. If you zoom in, [specifically] some edges ... [and] color blocks are quite distinct.''} (M8). At the data layer, participants use Exif metadata as a credibility cue to verify the capturing device and timestamps. P1 noted that \textit{``we examine the image's original data, checking whether the pictures are taken by the phone or by AIs as the evidence.''}
This shifts verification from evaluating image content to assessing the context of its production.

\textbf{\textit{Provenance cross-referencing.}} Merchants and platform workers search for images, or embed physical markers to establish the origin of evidence.
They used reverse image search to detect reused content. These methods operate on a temporal logic to \textit{``show that the image appeared earlier than the time they submitted it.''} (P1) Participants noted that a dedicated reverse searching database is currently under construction, and its coverage remains limited. Furthermore, this method is unevenly institutionalized and challenged by unique AI-generated inputs, where they cannot rely on the generation process to find matches (M3). To address this limitation, systems increasingly depend on physical anti-counterfeiting codes as material anchors (P5). 
When discrepancies arise, merchants can verify authenticity by scanning the code.

\subsubsection{Contextual and Evidentiary Verification}

\textbf{\textit{Temporal verification.}} Merchants ask for real-time, multi-angle videos as additional evidence, to increase synthesis difficulty and put additional burden of proof on attackers.
As M2 noted, \textit{``if the image looks abnormal ... we ask customer service to request ... a video.''}
However, M1 acknowledged the practical difficulty of enforcing real-time unboxing videos, where customers mostly declined.

\textit{\textbf{Logic verification.}} Adjudicators triangulate chat logs and logistical data to detect narrative contradictions. P1 described this as \textit{``a comprehensive review assessing whether there were promises ... or whether they were overtime situations.''} (P1) This audit identifies fraud when technically perfect visual evidence is contextually invalid, though it remains just a supplementary signal, which must also contradict the customer's specific claims to be effective.

\textit{\textbf{Reputation-based verification.}} Platforms adjust scrutiny levels based on customers' historical trustworthiness scores and behavioral patterns. Trusted accounts receive fast-path approvals, whereas flagged accounts undergo rigorous examination. M1 articulated this, \textit{``if you are a normal account, your first few refund requests are directly approved.''} Platform workers also recognize such tagging may have biases, where they may overly prioritize account history over case-specific forensics (P1).

\subsection{Intervention Criteria} 

\subsubsection{Platform-Level Criteria}


\textbf{\textit{Automated triage and exception handling.}} Platforms implement a conditional triage strategy that prioritizes automated resolution while designating high-risk claims, specifically refund-only requests, for mandatory human escalation. P4 noted that, \textit{``If it is refund only and no return, then [I think] in general it [should not] be handled automatically, as these claims signal potential abuse rather than standard product failures.''} The workflow follows a strict sequence where the system first attempts algorithmic resolution, and only when a customer rejects these proposals does the case route to a worker. P1 observed, \textit{``the AI provides some automatic responses ... but [if] they do not accept any of them ... subsequent judgments ... are all done by humans.''} 

\textbf{\textit{Weighted decision.}} Adjudication logic relies on multiple factors that weights monetary value, evidence consistency and user privilege tiers to determine the outcome. For low-value transactions, platforms accept certain losses  to minimize administrative costs. P1 mentions, \textit{``if it does not exceed 20 [RMB], then refund only appeals are basically directly approved ... which counts as internal loss.''} Conversely, for high-value products or flagged accounts, the system enforces strict checks, verifying if \textit{``the logic of the description is consistent with the evidence''} (P2). This stratified logic is further modulated by user status, such that accounts flagged for abnormal refund frequency triggered manual review to prevent their exploitation of rapid-settlement mechanisms. 

\textbf{\textit{Procedural heterogeneity and asymmetric appeals.}} We observed procedural heterogeneity across platforms, characterized by varying evidentiary standards. However, platforms similarly favor customers, and restrict merchant appeals. M10 contrasted one platform's strict protocol, which requires \textit{``videos or pictures, order numbers, or the outer box,''} with another platform's lenient model, where \textit{``you can tell them it is broken ... and the official customer service will refund you.''} Moreover, the merchant had fewer options if platforms already made a customer-favored decision.
P6 described \textit{``merchants cannot successfully appeal because the rules are weighted toward the consumer''}. Only those high-value cases have the chance of successful appeals, leaving merchants with limited defense strategies.

\subsubsection{Merchant-Level Criteria} 

Merchants consider three criteria when making decisions to handle disputes: store scale, customer history and plausibility.

\textbf{\textit{Store scale}} dictates the operational trade-off between managerial attention and dispute resolution capacity. Larger vendors often prioritize throughput over individual defense. M10 said, \textit{``we frequently become vulnerable to users who take advantage of loopholes because they are sometimes busy and do not have time to manage the store''} (M10). Conversely, while smaller merchants have the time for intensive verification, their defense is often constrained by the high procedural costs of gathering the comprehensive records required by platforms to validate their claims.

Merchants use \textbf{\textit{customer transaction history}} as a filter to identify high-risk customers and decide whether to refute a claim. This judgment process relies on behavioral metrics, such as purchase frequency, return rates, and address changes, beyond the evidence itself. M3 explained that they selectively challenge claims based on these heuristics, noting that \textit{``for customers that the platform is not likely to approve for refunds, I will push back.''} By restricting claims from accounts with frequent after-sale reports, merchants can mitigate abuse without the prohibitive cost of appealing each piece of evidence. 

\textit{\textbf{Physical plausibility}} checks based on product knowledge serve as a primary heuristic for detecting anomalies and intervention. Merchants evaluate whether claimed defects are consistent with an item's plausible properties. A toy merchant (M2) illustrated that \textit{``if the damage degree is very severe, I reject the claim because in normal use it would not produce this kind of problem.''} This acts as a heuristic, lowering merchants' cognitive load involved in screening artifacts.

\subsection{Resolution and Remediation Strategies}


\subsubsection{Negotiation and Direct Settlement} 

Merchants use direct communication, re-verification, and alternative remedies to resolve disputes and deter opportunistic claims. Immediate contact, such as phone calls via the platform, signals vigilance and serves as a compliance test. Buyers who evade these requests may be suspected of fabrication. M3 noted that, \textit{``we will ... ask them to provide proof, and they will say they already threw it away ...''} (M3). During this process, merchants use polite, administratively neutral tones, which helps in resolving disputes. P1 explained the objective of these interactions, \textit{``We are just soothing the user's emotions.''} (P1) 


Proposing physical returns or replacements allows merchants to distinguish genuine dissatisfaction from financial opportunism (M3). Fraudulent customers typically reject these additional requirements, prioritizing cash extraction. M14 observed, \textit{``You can do a return-and-refund or we can exchange it, but they do not return it, they want money.''} (M14) This behavioral filter identifies fraud by exposing attackers' refusal of non-monetary remediation.

Re-verification serves as an integrity test, as multi-view consistency remains difficult to synthesize across multiple formats. P10 remarked, \textit{``if you ask them to provide a few more pictures and videos, they usually cannot fabricate it.''} (P10) This process also ensures that documentation meets the standards required for third-party courier compensation (P2, M9).

\subsubsection{Escalation and External Recourse} 

Hierarchical escalation acts as an audit layer for authenticating high-risk evidence. For high-stakes products, individual merchants lack the authority to reject AI-fabricated claims independently, and such decisions require supervisory approval. P6 noted, \textit{``A worker can only determine that this evidence is invalid after reporting to the supervisor and after the supervisor also agrees.''} (P6) Complex or persistently unresolved disputes trigger specialized evaluation workflows involving dedicated experts or coordinators (P7). P5 explained, \textit{``If the issue is not solved, we may apply to the leader, and the leader will organize specialized people to evaluate it.''} (P5) 

Structured reporting mechanisms allow merchants to anchor allegations to specific transaction data points than general complaints. These context-sensitive interfaces help identify extortion or coordinated abuse. M3 described the workflow, \textit{``When you report through the image, it will automatically recognize all the images in that chat window, and you choose which image you want to report.''} (M3) These channels also facilitate batch fraud detection by correlating suspicious data, such as multiple accounts sharing identical shipping addresses (M5).

When standard adjudication channels fail, merchants leverage external channels, including media exposure and law enforcement, to bypass automated barriers. Publicity often supersedes standard decision logic. M2 reported, \textit{``Media intervention effectively pressured the platform to manually appealed again and returned the 50 [RMB] to the store ... when exploitation exceeds financial thresholds, we appeal to formal legal recourse.''} M5 articulated, \textit{``If abusive behavioral persists and the amount reaches a certain level, then you can go to the police.''}

\subsubsection{Adjustment and Loss Absorption} 

As described above, merchants and platform practitioners often mixed different verification strategies to facilitate dispute resolution (P5, M8). P5 stated, \textit{``these is definitively algorithmic involvement in the review process, but we also have human intervention. It is impossible to let the algorithm review everything while humans do nothing.''} (P5) P10 commented, \textit{``the system will generate steps and indicate the compensation range, and we compensate within that range.''} (P10) 

To minimize temporal and emotional burdens, stakeholders often internalize low-value dispute costs (M6-7). P1 noted, \textit{``if it does not exceed 20 CNY, some refund only applications will be approved directly ... This is our internal loss.''} (P1) Merchants also raise partial settlements to terminate conflicts. M5 detailed this tactic, \textit{``I express in a tactful way that the picture may be fake ... after his psychological defense is broken, I give him some compensation.''} (M5). Finally, merchants like M8 manage residual risks through selective disengagement. They chose not to respond to customers, and as M8 said, \textit{``customers did not continue further and just ignore it.''}

\section{RQ3: Constraints for Fraud Mitigation}

\subsection{Technical and Infrastructural Constraints}


\subsubsection{Advanced AI outpaces detection.}


Advancements in GenAI made fabricated evidence visually indistinguishable from authentic documentation. Participants observed that the high fidelity of synthetic images makes judgment difficult, and visual styles also vary across platforms. P3 illustrated, \textit{``after all kinds of AI-generated images increased ... I often cannot tell. At first glance, there seems to be no problem.''} With this evolution, even merchants are still learning new patterns, being able to reliably discern artifacts after being repeatedly deceived. As M5 remarked, \textit{``merchants might first believe it is real and proceed accordingly, and only after it happens again and again do you realize it was fake.''} (M5) 

Participants mentioned that current organizations lack automated verification tools to match the sophistication of AI manipulation. Consequently, authenticity judgments rely heavily on ad-hoc manual inspection and informal peer validation than technical detection (P6, M12). M12 emphasized this shift, \textit{``now it is not only handling the issue. You also have to judge what is real and what is fake. It really increases the workload.''} (M12) They must often cross-check with colleagues or use unreliable general AI tools to detect anomalies. P3 explained, \textit{``we do not have professional tools, and internally there is no AI verification tool ... Sometimes, especially this year, GenAI images are so deceptive that I cannot judge, so I ask colleagues to screen.''} Although merchants notice distinct surges in fraudulent claims following major promotional events, they still lack reliable detection mechanisms. For example, M2 commented, \textit{``we can only use AI to detect AI, but it always feel unconvincing.''}

GenAI also lowered the technical skill required for fabrication, enabling broader demographics to engage in opportunistic exploitation and increasing the workload for merchants. Unlike traditional editing software that requires expertise, prompt-based AIs allow users to generate deceptive evidence with minimal effort. M13 noted, \textit{``with AI tools, everyone can just give an instruction to generate an image. The user base expands, the cost of use drops, and more people get involved.''} M12 described this as a low-risk gamble for customers, \textit{``they are betting you can not tell ... if they win, they make money. If they lose, they still keep the item ... They have no cost.''}


Participants complained that platform's AI agents are restricted to text templates and cannot analyze visual evidence, leading users to bypass these barriers (P1). Despite being a paid service, these rudimentary bots frequently misinterpret cases, escalating dispute friction (M5, P8). P18 stated, \textit{``There may be such a function in theory, but right now the chatbot is still not smart, and we even have to pay for it every month.''} (M10) Even chatbots' keyword-based defenses are currently easy for attackers to bypass, forcing merchants to rely on manual heuristics (M10, M15). M15 observed, \textit{``Our detection tools mainly check watermarks, so once the watermark is removed, the tool cannot detect anything.''} (M15) Automated systems are therefore treated as secondary measures, \textit{``The platform has an AI-supported system too, but we judge manually first, and only use the system when we truly cannot tell.''} (M10) To address this, merchants demand standardized, on-page verification tools (P4, P10). P4 argued, \textit{``the platform should standardize this ... It should equip an AI image verification tool.''} (P4)  P10 further suggested, \textit{``it would be best if detection is available on the current chat page.''}


\subsubsection{Missing ground truth.}

Merchants face difficulties in verifying claims due to a lack of pre-shipment ground truth and documentation, particularly for perishables. As M3 explained, \textit{``Large furniture merchants or clothing merchants might have an initial shopping video, but for fresh produce and food we cannot do that, because during transport there are uncontrollable factors.''} (M3) 
Besides, merchants suspect AI forgeries but lack platform-verified tools for proving. M3 stated, \textit{``Right now we do not have any tool to prove that the customer used AI-generated images.''} (M3) Merchants therefore struggle to defend themselves when encountering suspected cases (M8). They also only had transaction histories of the customer within their storefront, lacking a lack of comprehensive data about customers, as opposed to platform workers who could have IDs, all transaction histories, and risk profiles of the customer.

Platform adjudication further disadvantages merchants by prioritizing speed and consumer retention over factual accuracy (P2).
M2 observed that \textit{``current automated reviews are optimized for rapid closure at the expense of verifying `image reality'.''} This model hinders merchants to reach platform staff to contest sophisticated fabrications (M2,P6).


Consequently, merchants like M8 advocated for a shift toward proactive risk signaling, suggesting that platforms \textit{``implement detection tools to identify tampered materials and issue warnings to offenders.''}
Besides, merchants proposed adjustments prioritizing auditability. P2 suggested transparent decision logic, \textit{``After desensitization, the basis for liability and the key evidence could be disclosed to both sides.''} (P2) M14 further advocated for system-level detection to flag potential fraud, \textit{``it could identify that an image has been edited later, or that there is suspicion of AI modification.''} (M14)



\subsubsection{Privacy and data fragmentation.}

10/29 participants thought that data privacy rules obstruct the development of cross-platform fraud databases. Although stakeholders value shared repositories for detecting GenAI fraud, they thought that disputes regarding data ownership and legal legitimacy stall collaboration. As P1 observed, \textit{``The data required for effective detection often raises questions regarding copyright infringement and jurisdictional scope.''} (P1) Therefore, the construction of these databases is a complex socio-legal challenge. Although P1 emphasized that sharing data across platforms is necessary to verify images, it remains an experimental idea with little implementation.

Limits in identity resolution hinders tracking attackers. P1 argued that even with real-name registration frameworks, linking disparate accounts to a single malicious actor remains difficult for them due to customers' device and network switching, which creates \textit{``a core design tension between enforcement efficacy and privacy compliance boundaries ... [as] detection mechanisms frequently break down when adversaries distribute fraudulent activity across distinct IP addresses and hardware, rendering indirect behavioral signals ineffective.''} 
%


To mitigate the fragmentation issues, participants advocated for layered architectures embedding verifiable signals into the supply chain (P5). P7 suggested a hybrid model, \textit{``we suggested using shared databases to automatically resolve clear-cut cases, such as web-scraped images, while resolving ambiguous instances for review.''} They also advocated for integrating automated detection with collective human oversight (P2), or leveraging multiple team-level reviews for complex cases. P4 added that platforms should use standard, outsourced detection reports to support verification.

They also proposed to differentiate attackers through profiles. M3 argued that platforms must \textit{``distinguish between compliant and non-compliant actors rather than applying generic policies''.} They noted that the current profiles rely on membership tier rather than risks (M3,M8,P1,P4), which should be improved in the future to correct the over-prioritization of product quality (M2), though it comes with privacy compliance challenges (P1). They also proposed to disclose liability rationales to both parties (M3).

\subsection{Procedural and Interactional Constraints}

\subsubsection{Structural bias and irreversibility.}


Adjudication processes lack actionable standards, forcing merchants and platform workers to balance platform rules, user satisfaction, and merchant rights without clear criteria (P2). 
P3 described AI suspicion labels as merely improvised delays, \textit{``we actually have no way to verify further. It is just a stalling tactic ... in training we were told to be vigilant, but there is no practical guidance on how to be vigilant, or how to identify it.''} 

Ambiguous adjudication typically favors buyers, converting suspected fraud into merchants' own losses (M2,P4,P21). Without formal GenAI policies, merchants face an asymmetric burden of proof, requiring extensive documentation to counter subjective buyer claims (M2, M8). Platforms granted refunds even when algorithms or merchants suspected manufactured defects. M2 stated, \textit{``the platform still supports a return or refund. We had no choice and could only treat it as daily loss.''} This policy imposes heavy financial burdens, including round-trip shipping and total resale value loss. P6 noted, \textit{``If the buyer has already used it and returns it under quality issues, the merchant cannot resell it, and the round-trip freight is a total loss.''} They explained, \textit{``only when we can pick out a problem will they let the case continue with more evidence. If we cannot find an issue, it is approved by default.''} (P4) 

The regulatory focus on throughput further limits merchants' ability to compel buyer cooperation or documentation (M5-6). They often face repetitive unverified allegations and risk reputational retaliation when enforcing evidence standards (M5, M9). M5 noted, \textit{``if they do not cooperate, they just insist that the item has a problem and refuse to admit it was caused by themselves.''} (M5) They complained that strict service metrics and automated platform interventions bypass their discretion (M9-10, M15). M10 marked that, \textit{``they may refund you after one conversation, and sometimes we do not even know it happened.''} (M10) M9 remarked, \textit{``they sometimes requires you to reply within minutes, and if you respond slowly, your score drops and you get restricted.''} (M9) 

They further highlighted that the administrative closure of disputes creates a state of irreversibility, preventing retroactive corrections even when physical inspection later exposes fraudulent manipulations (M5, M14). Regarding this lack of recourse, they noted, \textit{``if the platform has already paid the customer and the case is closed, and only then we find problems with the item, it is very hard to get the money back.''} (M14) 

Participants demanded explicit guidelines and integrated workflows connecting to risk management teams. P4 emphasized the need for authoritative boundaries, \textit{``give tools, but also give standards. We need a real boundary and clear example cases ... so we can follow an official standard and not make mistakes.''} (P4) They also wanted the procedural rights to challenge claims with formal tools, \textit{``we need an AI detection tool, and we need the right to question or reject.''} (P1) Finally, M14 argued for a shift from credit-based to evidence-based judgment, \textit{``you should not use platform data about credit to judge the quality of the returned goods.''} (M14)



\subsubsection{Bottlenecks and automation bias.}

Massive ticket volumes and speed-based incentives compel merchants to prioritize throughput (P3, P8, P10). \textit{``We handle so many tickets every day that we simply do not have time to analyze what is going on.''} (P8) Financial incentive mismatches further discourage investigation, as they mostly prioritize response rates, \textit{``we also have a response rate requirement, otherwise you lose the performance bonus.''} (P10) Automated systems often misclassify forgeries as standard quality issues, depriving merchants of forensic context (M8, P8). P8 noted, \textit{``the robot cannot recognize it. It might treat an AI-generated image as a quality issue.''} (P8) 

These workflows sometimes even settle disputes automatically with minimal evidence (P3). \textit{``It goes through the platform's automated chain and gets refunded with just two pictures.''} (P3) Premature intervention hinders negotiation and facilitates fraud (M10, P10). \textit{``Sometimes you have not even exchanged a few messages and the platform automatically steps in.''} (M10) 

Consequently, merchants requested embedded authenticity tools, \textit{``it would be faster if, on the current chat page, there were a function like image screening or authenticity checking.''} (P10) P1 also highlighted the urgent need for dedicated reporting pathways, noting that \textit{``we can only make the judgment ourselves.''} (P1)


\subsubsection{Suppressed merchant discourse.}

Negotiation frequently fails during the initial fact-finding stage, as customer are often reluctant to cooperate, and merchants can hardly raise compelling evidence. This creates a deadlock where merchants can neither verify claims nor persuade customers to alter their stance (M9, M12). M5 described, \textit{``If they do not cooperate, they just insist that the item has a problem.''} Merchants also lack the authority to compel evidence collection, as M6 noted \textit{``some people say they did not take any photos, and you cannot force them to take pictures after they receive the package.''} (M6)

Platform algorithms limit merchant expression, preventing them from contesting claims. Merchants report that these systems aggressively flag negativity or sarcasm, forcing them to neutral responses. M12 explained that because even subtle negativity is penalized, many have resorted to using generic emojis, such as a ``smiley face'', to signal frustration. They also self-censored messages before posting to mitigate risk of being flagged by platforms (M12).

The threat of reputational damage and algorithmic ranking penalties compels merchants to accept losses. As M6 described, \textit{``In an era of abundant supply, sellers are always in a weak position ... once your sales drop, your ranking will drop.''} Consequently, merchants compensated whenever it is cheaper than managing the fallout of a bad rating. M14 said, \textit{``The cost of getting a bad review is higher than selling several pieces of clothing ... if you do not pay, they will definitively give you a bad review.''} 




\subsection{Economic and Resource Constraints}

\subsubsection{Economic imbalance in proof production.}
12/17 merchant-side participants reported that listing evidences imposes disproportionate costs on merchants, compared to the minimal effort for customer-initiated refunds. M3 highlighted this asymmetry, \textit{``[once] a quality issue is claimed ... the platform [issues a] refund. In many cases, there is no proof at all. Just a single sentence can trigger intervention ... [Yet] for us, the profit on one fruit order is about one CNY. If I report one customer to the market regulator ... the labor cost is simply too high.''} (M3) Therefore, this economic imbalance, compounded by the persistently low appeal success rates, deters merchants from investing labor for low-value transactions.

Resolution also requires extensive evidence across the entire product lifecycle, including continuous monitoring. M2 emphasized the hardness of this strong proof, \textit{``To succeed you need to prepare complete supporting materials. We have surveillance for every shipment ... only with very strong evidence will they agree to reject the claim.''} (M2) Even so, successful ones still face long processing cycles that delay financial reimbursement (M13).

\subsubsection{Reputational risk and incentive misalignment.}
11/17 merchant-side participants also put efforts to manage reputational risk and avoid platform penalties. They frequently pay compensation to prevent negative ratings or prolonged negotiations, treating these payments as a cost of doing business. P1 explained this, \textit{``some sellers will compromise to avoid any later disputes ... They are afraid of bad reviews.''} M13 emphasized the economic trade-off, \textit{``a single negative review costs more than selling several items. If they want ten CNY, you can just give.''}
Platform or merchant-side services also do not feature rewards for identifying fraud, which discourages them from investing effort in adjudication. P1 noted, \textit{``Even if you identify [fraud] and stop the loss ... [there] is no reward mechanism ... so there is no motivation.''} 

\section{Discussion}

\subsection{Technical Feasibility of Resolutions}

\begin{table*}[t]
\centering
\caption{Technical strength and workflow constraints of candidate defenses.}
\label{tab:defense-gaps}
\small

\begin{tabularx}{\textwidth}{@{}
P{0.23\textwidth}
P{0.23\textwidth}
Y
P{0.13\textwidth}
@{}}
\toprule
\textbf{Resolution Method} &
\textbf{Theoretical Strength} &
\textbf{Workflow Constraint} &
\textbf{Linked Challenge} \\
\midrule

\multicolumn{4}{l}{\textit{\textbf{Manual Inspection \& Heuristic Judgment}}} \\

Physics consistency checks &
Identifies visual artifacts~\cite{joslin2024double} &
Fails against AI-generated defects~\cite{lu2023seeing} &
\icons{\tech} \\

Adversarial spatial interrogation &
Discriminative features~\cite{gu2021spatiotemporal,zhang2024deepfake} &
High user friction; hard-to-require evidence &
\icons{\struct\ \inter} \\

Collaborative consensus &
High consensus accuracy~\cite{bitaab2023beyond,zhang2021fraud,shehnepoor2021dfraud3} &
Massive participation needed; platform data silos &
\icons{\datafrag\ \struct} \\

\midrule
\multicolumn{4}{l}{\textit{\textbf{Technical \& Automated Authentication}}} \\

Algorithmic risk screening &
High precision~\cite{zhu2021dissecting,viswanath2014towards,deblasio2017exploring} &
Transient accounts; short refund windows &
\icons{\struct\ \datafrag} \\

Digital forensic inspection &
Standard historical protocol~\cite{zhu2025neural} &
Easily spoofed; compressed metadata &
\icons{\tech\ \struct} \\

\midrule
\multicolumn{4}{l}{\textit{\textbf{Contextual \& Evidentiary Verification}}} \\

Temporal verification &
Strong theoretical defense~\cite{tursman2020towards,zhang2023understanding} &
Hardware circumvention; hard-to-mandate live capture &
\icons{\tech\ \inter} \\

Reputation-based verification &
Sybil resilient~\cite{kumar2018rev2,xu2017online,mazloomzadeh2024reputation} &
Gaming vulnerability; account switching &
\icons{\struct\ \datafrag} \\

\bottomrule
\end{tabularx}

\vspace{2pt}
\begin{minipage}{0.98\textwidth}
\footnotesize
\textit{Notes.}
\tech~= technical complexity;
\struct~= structural/platform barriers;
\econ~= economic imbalance;
\inter~= interactional constraints;
\datafrag~= data fragmentation.
\end{minipage}
\end{table*}

We discussed the technical feasibility of various resolution mechanisms, as summarized in Table~\ref{tab:defense-gaps}. While existing literature shows high theoretical performance for some mitigation~\cite{zhu2021dissecting,viswanath2014towards,bitaab2023beyond,tursman2020towards}, we found that their practical efficacy is limited by multiple constraints, such as cognitive limitations, platform business logic, and evolving adversarial manipulations.

\textbf{Algorithmic and reputation-based screening.} Algorithmic risk screening exhibits high theoretical efficacy in controlled environments. For instance, Zhu et al.~\cite{zhu2021dissecting} and Viswanath et al.~\cite{viswanath2014towards} both achieved over 94\% anomaly detection rate. However, our findings indicate that the operational pace of e-commerce platforms limits deployment effectiveness. Fraudulent accounts operate in short temporal windows. Prior measurements show that 35\% of fraudulent accounts are terminated prematurely and 90\% are closed within four days~\cite{deblasio2017exploring}. Therefore, algorithmic intervention may arrive too late to intercept instant refund workflows. Furthermore, while past work reports that reputation-based trust scoring can achieve up to 84.6\% accuracy~\cite{kumar2018rev2}, it suffers from severe precision degradation, falling to 60\%--80\%~\cite{mcdonald2019reliability} or exhibiting recall as low as 0.65~\cite{xu2017online} when facing sophisticated, organized frauds. This unreliability may force platforms to fall back on manual review, reducing algorithmic defenses' scalability.

\textbf{Physics-consistency and spatial interrogation.} Our analysis confirms the ineffectiveness of human-initiated checks. As GenAI advances, humans fall to near-random guessing, with average detection accuracy at 61.34\%~\cite{lu2023seeing}. Although humans may marginally outperform baselines when identifying artifacts in peripheral regions~\cite{joslin2024double}, this is insufficient for enterprise-grade security. To compensate, spatial interrogation leverages multi-angle analysis, exploiting AI models' inability to render flawless physical geometries. These algorithms achieve robust theoretical efficacy, reaching 91.78\% accuracy~\cite{gu2021spatiotemporal} and 65\% frame-level performance across major deepfake datasets~\cite{zhang2024deepfake}. However, our interviews indicate that deploying spatial interrogation in e-commerce faces severe user friction. As platforms prioritize consumer retention, they can hardly enforce these standards.
Besides, many merchants' overhead of shipping, sorting, and manual inspection exceeds the residual value of products. On the contrary, as GenAI reduces the attack cost to near zero, attackers could easily carry out the attack, causing merchants to compromise totally.

\textbf{Collaborative consensus and decentralized verification.} Multi-node verification, label aggregation and collaborative consensus reach up to 98.34\% accuracy~\cite{bitaab2023beyond,zhang2021fraud}, while heterogeneous information networks achieve F1 scores of 0.666 to 0.70 on crowdsourced datasets~\cite{shehnepoor2021dfraud3}. Despite these algorithms, applying decentralized verification to refund fraud is constrained by data fragmentation and privacy boundaries. Merchants emphasized that strict data fragmentation prevent the formation of cross-platform fraud database. Therefore, merchants remain isolated, lacking the massive, aggregated participation for consensus models against coordinated synthetic evidence attacks. 


\textbf{Forensic inspection.} Traditional digital forensics, such as Exif metadata analysis, have become increasingly easy to bypass due to the lack of cryptographic binding. Sophisticated adversaries now use neural networks to actively remove AI artifacts from generated media, successfully bypassing basic image signal processing defenses~\cite{zhu2025neural}. Temporal verification (e.g., real-time liveness checks) counters static image manipulation, offering accuracy of 0.60~\cite{tursman2020towards}. Yet, as highlighted by recent system-level security analyses~\cite{zhang2023understanding}, fraudsters actively circumvent these temporal checks via hardware-level exploits, using camera hijacking and OS-level video injection tools. As we found merchants already struggle to find flaws from uncooperative buyers, mandating hardware-secured, real-time video streams in disputes remains difficult.

\subsection{Generalizability of GenAI-enabled Fraud}
We discussed the generalizability of GenAI-enabled frauds through reviewing refund policies across representative global e-commerce platforms. As shown in Table~\ref{tab:refund_comparison}, we found many platforms feature refund only policies, which are trigged by simple heuristics like price. These refund mechanisms, originally designed to minimize reverse logistic costs for inexpensive goods, may expand the attack surface for GenAI-enabled frauds. However, susceptibility may vary across regions due to differences in transaction volume, price, and storefront structures. For instance, high-frequency, low-value models (e.g., Douyin, Temu) may be more prone to automated exploitation, whereas the low-frequency EU market currently exhibits lower fraud intensity. However, the global surge in reported refund frauds~\cite{George2024RefundAbuse,Perry2026ImageBasedFraud} underscores that GenAI-enabled fraud represents a pervasive threat to the integrity of global digital commerce. Notably, mature platforms such as Amazon~\cite{amazon_refund_fraud_2025} have responded by deploying automated countermeasures to mitigate these evolving refund frauds.

\begin{table}[htbp]
\centering
\caption{Comparison of global e-commerce refund mechanisms.  \textbf{Refund Only}: full refund, no return. \textbf{Partial Refund}: partial refund, item kept. \textbf{Return \& Ref}: refund after return. \textbf{Refund Criteria}: logic for refund. Refer to Appendix~\ref{app:comparison} for detailed policy and criteria explanation.}
\label{tab:refund_comparison}
\resizebox{0.5\textwidth}{!}{
\begin{tabular}{lccccccc}
\hline
\textbf{Platform} & \makecell[c]{\textbf{Primary} \\ \textbf{Region}} & \makecell[c]{\textbf{Refund} \\ \textbf{Only}} & \makecell[c]{\textbf{Partial} \\ \textbf{Refund}} & \makecell[c]{\textbf{Return} \\ \textbf{\& Ref}} & \textbf{Refund Criteria} \\ \hline

\textbf{Temu} & World & $\checkmark$ & $\checkmark$ & $\checkmark$ & -- \\ 
\textbf{TikTok Shop} & World & $\checkmark$ & $\checkmark$ & $\checkmark$ & Price \\ 
\textbf{Shopee} & Asia & $\checkmark$ & $\checkmark$ & $\checkmark$ & Price \\ 
\textbf{Lazada} & Asia & $\checkmark$ & $\checkmark$ & $\checkmark$ & -- \\ 
\textbf{AliExpress} & World & $\checkmark$ & $\checkmark$ & $\checkmark$ & Price \\ 
\textbf{Shopify} & World & $\checkmark$ & $\checkmark$ & $\checkmark$ & -- \\ 
\textbf{Amazon} & World & $\checkmark$ & & $\checkmark$ & Price \\ 
\textbf{eBay} & World & $\checkmark$ & $\checkmark$ & $\checkmark$ & -- \\ 
\textbf{Walmart} & USA & $\checkmark$ & $\checkmark$ & $\checkmark$ & -- \\ 
\midrule
\textbf{Taobao} & China & $\checkmark$ & $\checkmark$ & $\checkmark$ & Credit \\ 
\textbf{JingDong} & China & $\checkmark$ & $\checkmark$ & $\checkmark$ & Credit \\ 
\textbf{Douyin} & China & $\checkmark$ & $\checkmark$ & $\checkmark$ & Price \\ 
\textbf{Dewu} & China & $\checkmark$ & $\checkmark$ & $\checkmark$ & Credit \\ 
\textbf{Meituan} & China & $\checkmark$ & $\checkmark$ & & Property \\ 
\toprule
\end{tabular}
}
\end{table}

\subsection{Implications}

$\bullet$ \textbf{\textit{Facilitating cross-platform collaborative defense.}} Our findings suggest that many platforms are currently concurrently exploited by malicious actors to scale fraudulent activities. Future e-commerce systems should implement privacy-preserving collaborative defenses to disrupt these patterns. Instead of sharing raw personal data, platforms could exchange anonymized behavioral signals or hashed evidence fingerprints~\cite{zhang2024adanonymizer}. Such a shared reputation repository would enable merchants and platforms to identify recycled evidence or repeat offenders across different services without compromising regulatory compliance or user privacy.

$\bullet$ \textbf{\textit{Bridging and digital-physical gap via verifiable anchors.}} To mitigate the risks of GenAI-fabricated evidence, developers should incorporate physical or digital anchors, such as watermarks, into the logistics chain. We suggest embedding verifiable, tamper-evident markers, such as randomized patterns or unique identifiers during product packaging. As these markers are difficult to simulate or edit without prior knowledge, they serve as grounding elements for submitted evidence during the refund phase. Coupling these with shipment-specific QR codes ensures that digital evidence remains strictly tied to a unique physical transaction, effectively raising the cost of large-scale evidence fabrication.

$\bullet$ \textbf{\textit{Implementing risk-adaptive adjudication workflows.}} To address the tension between operational efficiency and fraud detection, platforms should move away from rule-based automation toward risk adaptive workflows, similar to those on Amazon~\cite{amazon_refund_fraud_2025}. Platforms could use a tiered architecture where flagged accounts are managed via standard rejection algorithms, while ambiguous cases are automatically routed to specialized human oversight with dedicated, explainable tools~\cite{zhang2025exploring}. By doing so, these stakeholders could shift the burden of countering complicated AI fraud.

\section{Limitations}

Our paper is subject to several limitations. First, the semi-structured interview format allow for a rich description of emerging threat vectors, but the sample size and format do not support statistical generalization regarding the prevalence of GenAI fraud across the entire e-commerce ecosystem. Second, our investigation focuses exclusively on the Chinese e-commerce market (e.g., Taobao, Pinduoduo, Douyin). While we found substantial risks on these platforms, and Western platforms also have refund fraud risks\footnote{https://pages.ravelin.com/agentic-commerce-fraud-report/}, specific findings may not directly transfer to Western platforms with different policy and infrastructures. Third, our data relies on self-reported experiences from merchants and platform workers, with no attackers involved due to ethical concerns. We also did not conduct real-world empirical measurements. While we triangulated these perspectives, and discussed these findings regarding feasibility, retrospective reports are subject to recall bias, and may not fully capture each attack, especially those that merchants fail to recognize. Finally, the advancement of GenAI may limit the applicability of specific attack vectors. However, the underlying vulnerabilities originating from platform verification are likely to persist.

\section{Conclusion}

This paper investigates refund fraud, especially GenAI-enabled frauds, through interviewing merchants and platform workers. We first identify refund fraud attack vectors across four transaction lifecycle stages. Second, we reveal that while merchants rely on a combination of general AI tools and manual heuristics to verify manipulated artifacts, the rapid evolution of GenAI increasingly undermines these detection efforts. The negligible cost of GenAI-enabled fraud also imposes an asymmetric burden of proof on merchants. We further categorize the barriers to fraud resolution into technical, procedural, and economic aspects. We finally propose implications emphasizing collaborative database sharing, physical-digital grounding, and risk adaptive adjudication workflows.

\begin{acks}
We acknowledge the use of Gemini 3.1 Pro and ChatGPT strictly for minor editing, specifically grammar and style polishing. Authors retain full responsibility for the accuracy, originality, and integrity of this paper.
\end{acks}


\bibliographystyle{ACM-Reference-Format}
\bibliography{main}

\appendix

\section{Ethical Considerations}

We acknowledged potential ethical concerns and considered these implications throughout the interviews. Our research was conducted with the approval of our institution's Institutional Review Board (IRB). We detailed our considerations, in accordance to the principles outlined in the Belmont Report~\cite{beauchamp2008belmont} and the Menlo Report~\cite{bailey2012menlo}.

\textbf{\textit{Respect for persons:}} We obtained informed consent from all participants prior to interviews. Participants were informed that they could freely quit the interview, and require the deletion of their data, at any time before, during or after the interview, without any reasons. Given the sensitive nature of the topic, where merchants and platform workers disclosed vulnerabilities and internal technical gaps, we strictly anonymized all personal and institutional identifiers. Participants were informed of their right to withdraw from the study at any time without penalty. We analyzed the potential socio-technical impact ofour interviews on the following groups.

$\bullet$ Impact on merchants and platform workers: As merchants disclosed internal technical gaps and financial vulnerabilities related to fraud, we prioritized confidentiality to prevent competitive disadvantage or retaliatory attacks. We address the power asymmetry between platforms and merchants by highlighting the asymmetric burden of proof and the economic challenge caused by refund only policies.

$\bullet$ Impact on consumers: To protect the legitimate consumer interest in efficient dispute resolution, our recommendations emphasize risk-adaptive adjudication rather than total restrictions. This ensures that while fraud is mitigated, the friction for honest users remain low.

$\bullet$ Impact on researchers: To protect the researchers, we refrained from directly contacting attackers and malicious groups. To protect the broad research community and the public from the dual-use of our findings, we abstracted technical exploits (e.g., models) and redacted sensitive parts. By doing so, we ensure that our paper does not serve as a guideline for fraudsters.

\textbf{\textit{Beneficence:}} We prioritized the minimization of harm to participants and the broad ecosystem. To prevent the dual-use of our findings, where detailed descriptions of fraud might serve as an instructional guide for attackers, we have abstracted specific technical exploits and redacted sensitive keywords or prompts used by fraudsters. The research aims to benefit the community by exposing systemic vulnerabilities and proposing robust defense mechanisms against AI-enabled fraud.

\textbf{\textit{Justice:}} The burdens and benefits of this research were distributed equitably. We recruited a diverse range of stakeholders, including merchants and platform workers who are often disproportionately affected by fraud but underrepresented in policy design. Our recommendations aim to rectify the power asymmetry that currently disadvantages these groups.

\textbf{\textit{Respect for law and public interest:}} Our investigation was conducted in compliance with relevant data protection regulations and platform terms of service. We focused on analyzing the mechanisms of fraud to serve the public interest of maintaining a secure and trustworthy of digital economy, rather than facilitating or encouraging valid policy violations.






\section{Interview Scripts}\label{app:interview_script}

\subsection{Interview Script for Platform Workers}

\textbf{Part 1: Professional background and general workflow}

1. Could you briefly describe your role, your experience level, and the types of disputes you typically handle?

2. What are the primary performance metrics or key performance indicators (KPIs) that guide your daily workflow?

\textbf{Part 2: GenAI threat landscape and encounters}

3. Can you describe a recent dispute where you suspected a buyer was using AI or automated software, whether to generate a fake persona, fabricate evidence, or manipulate the platform's logistics?

4. Can you walk me through a case where you strongly suspected the evidence was fabricated, but you ultimately could not definitely prove it or had to approve the refund anyway? What made the evidence difficult to verify?

\textbf{Part 3: Verification practices}

5. When you suspect fraudulent evidence, what is your standard verification process?

6. How do you interact with the platform's risk-screening tools and resolution process?

7. Can you share an instance where you felt the platform's automated systems misclassified a case? How did you handle that discrepancy?

8. (If applicable) What are your knowledge source for these verification or tools used?

\textbf{Part 4: Challenges}

9. In your experience, what are the primary challenges you face when encountering fraudulent activities?

10. Are there rules that you feel constrain your ability to investigate refund fraud fraudsters? Why?

\textbf{Part 5: Countermeasures}

11. What specific tools, shared databases, or verification standards would help your dispute resolution process?

12. Is there anything else you would like to share about around refund fraud?





\subsection{Interview Script for Merchants}

\textbf{Part 1. Professional background and general workflow}

1. Could you briefly describe your experience managing your store and your general professional background in e-commerce?

2. What is your standard workflow when you receive and process refund requests from customers?

\textbf{Part 2. Specific cases of synthetic evidence}

3. Can you share a recent case where you felt a buyer submitted AI-edited or synthetically altered evidence to justify a refund?

4. What was your initial reaction upon seeing the request and the provided evidence?

5. How did you go about identifying the fraud? What specific details, anomalies, or tools did you rely upon to determine the evidence was likely synthetic?

6. Could you walk me through the reasoning process that led you to that conclusion?

\textbf{Part 3. Decision-making logic and operational practices}

7. In situations where you suspect the evidence is synthetic, what is your logic for deciding whether to immediately issue the refund, or to reject the request and initiate an appeal?

8. How do you balance these decisions within your daily operational practices, considering different factors?

\textbf{Part 4. Challenges and expectations}

9. What are the primary challenges or limitations you encounter?

10. What specific countermeasures do you expect or hope to see developed in the future to protect merchants?

11. Is there anything else you would like to add regarding your experiences and refund fraud?


\section{Participants' Demographics}

Table~\ref{tab:merchants} and~\ref{tab:platforms} showed the merchants and platform workers' demographics separately.

\begin{table}[htbp]
\centering
\caption{Participants' demographics (merchants group).}
\label{tab:merchants}
\resizebox{\linewidth}{!}{%
\begin{tabular}{clllll}
\toprule
\textbf{ID} & \textbf{Category} & \textbf{Platform} & \textbf{Experience} & \textbf{Age} & \textbf{Gender} \\ \midrule
1  & Electronic Products    & -                 & 5 Years & - & M \\ 
2  & Toy Store              & -                 & 4 Years & - & F \\ 
3  & Fruit Merchant         & -                 & 8 Years & - & F \\ 
4  & Cosmetics              & Taobao \& Douyin  & 4 Years & 18-25 & F \\ 
5  & Women's Clothing       & Taobao            & 2 Years & 26-35 & M \\ 
6  & Sports Equipment       & Amazon            & 7 Years & 36-45 & M \\ 
7 & Furniture              & Taobao            & 1 Year  & - & M \\ 
8 & Aromatherapy           & Taobao            & 2 Years  & - & F \\ 
9 & Cosmetics              & Taobao            & 2 Years & - & F \\ 
10 & Makeup Brush Store & Pinduoduo & 2 Years & 18-25 & F \\ 
11 & Jewelry \& Hair Accessories & Pinduoduo \& Douyin \& Redbook & 12 Years & 26-35 & F \\
12 & Bags, Beeding, DIY Materials & Xianyu \& Taobao \& Douyin & 4 Years & - & F \\
13 & Men's Wear \& Fresh Produce & Taobao \& JD \& Pinduoduo & 10 Years & 26-35 & F \\
14 & Men's Wear \& Daily Necessities & Taobao \& JD \& Pinduoduo & 7 Years & 26-35 & M \\
15 & General Merchandise & Taobao & 3 Year & 18-25 & M \\
16 & Clothing & Pinduoduo & 2 Year & - & M \\
17 & Clothing \& Cosmetics & Tmall Global & 2 Year & - & F \\
\bottomrule
\end{tabular}
}%
\end{table}

\begin{table*}[t]
\centering
\caption{Platform-side participants' role coverage in refund workflows.}
\label{tab:platforms}
\scriptsize
\begin{tabularx}{\textwidth}{@{}
p{0.05\textwidth}
p{0.25\textwidth}
p{0.12\textwidth}
p{0.32\textwidth}
p{0.20\textwidth}
@{}}
\toprule
\textbf{ID} &
\textbf{Role Type} &
\textbf{Platform} &
\textbf{Workflow Perspective} &
\textbf{Operational Scope} \\
\midrule
P1  & After-sales (Outsourced)            & Taobao    & Standardized after-sales handling & Rule-bound execution / escalation \\
P2  & After-sales (Outsourced)            & Pinduoduo & Standardized after-sales handling & Rule-bound execution / escalation \\
P3  & After-sales (Outsourced)            & Meituan   & Standardized after-sales handling & Rule-bound execution / escalation \\
P4  & After-sales Service                 & Dewu      & Case-level return/refund processing & Case handling under platform rules \\
P5  & Customer Service                    & Pinduoduo & User-facing dispute communication & Complaint intake / limited authority \\
P6  & Internal After-sales                & Pinduoduo & Internal procedure coordination & Broader procedural visibility \\
P7  & Official Service (Beauty)           & Taobao    & Platform service and category disputes & Rule application in assigned scope \\
P8  & Official Service                    & Pinduoduo & Platform service and dispute handling & Rule application in assigned scope \\
P9  & Customer Service                    & Pinduoduo & User-facing dispute communication & Complaint intake / limited authority \\
P10 & After-sales (Outsourced, Footwear)  & Dewu      & Standardized after-sales handling & Rule-bound execution / escalation \\
P11 & Official Customer Service           & Pinduoduo & Platform service and dispute handling & Rule application in assigned scope \\
P12 & Technical Staff                     & Pinduoduo & Algorithmic risk operations & System-level support \\
\bottomrule
\end{tabularx}

\vspace{2pt}
\begin{minipage}{0.98\textwidth}
\footnotesize
\textit{Note.} Role type and workflow coverage are summarized from participants' self-reported positions. 
\end{minipage}
\end{table*}

\section{Qualitative Codebook}\label{sec:codebook}

Based on the interview data, we presented a qualitative codebook to reflect the taxonomy of threat vectors, verification mechanisms, and systemic constraints, as shown in Table~\ref{tab:codebook}.

\begin{table*}[htbp]
\centering
\caption{Qualitative codebook for GenAI-enabled refund fraud analysis.}
\label{tab:codebook}
\small
\begin{tabular}{@{} l p{4.5cm} p{8cm} @{}}
\toprule
\textbf{Theme} & \textbf{Code} & \textbf{Description} \\
\midrule
\multirow{9}{*}{\shortstack[l]{\textbf{Threat Vectors} \\ (RQ1)}} 
& \textbf{Automated promotion and subsidy arbitrage} & Exploiting promotional algorithms and shipping subsidies using coordinated bulk accounts. \\
& \textbf{Synthetic persona construction} & Utilizing GenAI to fabricate human-like behavioral histories and profiles to evade detection. \\
& \textbf{Entry barrier reduction} & Lowering the technical expertise required for systematic fraud through accessible AI generation tools. \\
& \textbf{Hyper-realistic defect fabrication} & Synthesizing physical damage that respects material properties and physical logic to deceive visual inspection. \\
& \textbf{Contextual attribute forgery and digital recycling} & Forging communication records and reusing identical synthetic evidence across multiple stores simultaneously. \\
& \textbf{Algorithmic bypass and asynchronous substitution} & Exploiting automated chatbots to secure immediate refunds before merchants can physically inspect returns. \\
& \textbf{Carrier-side interception} & Rerouting shipments mid-transit after prematurely securing an automated platform refund. \\
& \textbf{Off-platform double extortion and entrapment} & Weaponizing fabricated content publicly or migrating to external channels to extract redundant refunds. \\
& \textbf{Regulatory exploitation and psychological coercion} & Leveraging administrative complaints and emotional pressure to force merchants into financial settlements. \\
\midrule
\multirow{8}{*}{\shortstack[l]{\textbf{Verification \&} \\ \textbf{Resolution} \\ (RQ2)}} 
& \textbf{Manual inspection and heuristic judgment} & Evaluating evidence authenticity through physics consistency, spatial interrogation, and collaborative peer consensus. \\
& \textbf{Technical and automated authentication} & Deploying algorithmic risk screening, metadata forensics, and cross-referencing provenance markers. \\
& \textbf{Contextual and evidentiary verification} & Cross-referencing logical narratives, temporal data (e.g., live video), and historical account reputations. \\
& \textbf{Platform-level criteria} & Adjudicating disputes via automated triage, weighted decision metrics, and asymmetric procedural rules. \\
& \textbf{Merchant-level criteria} & Filtering and addressing claims based on operational scale, historical transaction data, and physical plausibility. \\
& \textbf{Negotiation and direct settlement} & Requesting physical returns, multi-angle proof, or alternative remedies to distinguish authentic dissatisfaction from fraud. \\
& \textbf{Escalation and external recourse} & Triggering supervisory reviews, specialized data reporting, or law enforcement intervention for complex disputes. \\
& \textbf{Adjustment and loss absorption} & Internalizing low-value dispute costs or executing partial compensation to minimize operational and emotional friction. \\
\midrule
\multirow{8}{*}{\shortstack[l]{\textbf{Systemic} \\ \textbf{Constraints} \\ (RQ3)}} 
& \textbf{Advanced AI outpaces detection} & The visual fidelity of synthetic evidence exceeding the capabilities of current manual heuristics and detection tools. \\
& \textbf{Missing ground truth} & The structural inability to establish pre-shipment documentation, particularly limiting defenses for perishable goods. \\
& \textbf{Privacy and data fragmentation} & Legal and compliance boundaries preventing the establishment of cross-platform intelligence sharing and centralized registries. \\
& \textbf{Structural bias and irreversibility} & Platform adjudication frameworks that structurally favor consumers and lack mechanisms for retroactive correction. \\
& \textbf{Bottlenecks and automation bias} & High dispute volumes forcing reliance on automated systems that frequently misclassify forged artifacts as genuine quality issues. \\
& \textbf{Suppressed merchant discourse} & Algorithmic moderation restricting merchants' ability to actively contest claims or compel evidence collection. \\
& \textbf{Economic imbalance in proof production} & The disproportionate labor and procedural cost required for merchants to disprove near-zero cost synthetic fraud. \\
& \textbf{Reputational risk and incentive misalignment} & The continuous threat of negative ratings coercing merchants into accepting fraudulent claims as a standard operational cost. \\
\bottomrule
\end{tabular}
\end{table*}

\section{Comparison of Platform Policy Links}\label{app:comparison}

We detail the platform policy links analyzed in this paper. In Table~\ref{tab:refund_comparison}, platforms that explicitly disclose refund criteria are identified in the ``Price'' column. Undisclosed criteria are denoted by ``--''. Based on our interviews, we categorized the refund criteria of Chinese platforms into three primary types: price-based (denoted as ``Price''), customer credit-based (denoted as ``Credit''), and product-specific such as perishable goods (denoted as ``Property''). Note that neither the list of criteria nor the selection of e-commerce platforms is intended to be comprehensive.

\begin{description}
    \item[Temu:] \url{https://www.temu.com/return-policy.html}
    \item[TikTok Shop:] \url{https://seller-us.tiktok.com/university/essay?knowledge_id=984433248438058}
    \item[Shopee:] \url{https://help.shopee.com.my/portal/4/article/77221-Refunds-and-Return-Policy}
    \item[Lazada:] \url{https://sellercenter.lazada.sg/helpcenter/s/faq/knowledge?categoryId=1000028206&m_station=BuyerHelp&questionId=1000150835}
    \item[AliExpress:] \url{https://customerservice.aliexpress.com/category?id=21013751}
    \item[Shopify:] \url{https://help.shopify.com/en/manual/fulfillment/managing-orders/returns/return-rules}
    \item[Amazon:] \url{https://sellercentral.amazon.com/help/hub/reference/external/G3RZC8DVQCDTCQ3B}
    \item[eBay:] \url{https://www.ebay.com/help/selling#returns-refunds}
    \item[Walmart:] \url{https://www.walmart.com/help/article/walmart-marketplace-return-policy/63c3566a9d3546858582acae2fbfdb7e}
    \item[Taobao:] \url{https://world.taobao.com/lang/en-us/shopping-guide/1899321028157374464.htm}
     \item[JingDong:] \url{https://help.jd.com/epthelp/question-639.html}
    \item[Douyin:] \url{https://lifexue.com/knowledge/detail/123777}
    \item[Dewu:] \url{https://agreement.dewu.com/client/view?id=127AG56CG1DDA&from=5f83ff7b57c2c0780741fd42}
    \item[Meituan:] \url{https://rules-center.meituan.com/rule-detail/132/5}
\end{description}

\end{document}